\documentclass[preprint]{ptephy}

\preprintnumber{KYUSHU-HET-142}

\usepackage{amssymb}
\usepackage{amsthm}
\usepackage{amsmath}
\usepackage{booktabs}
\DeclareMathOperator{\tr}{tr}

\newcommand{\Slash}[1]{{\ooalign{\hfil/\hfil\crcr$#1$}}}
\numberwithin{equation}{section}

\usepackage{url}

\begin{document}

\title{Lattice energy--momentum tensor from the Yang--Mills gradient flow---%
a simpler prescription---}

\author{\name{\fname{Hiroki} \surname{Makino}}{1},
\name{\fname{Hiroshi} \surname{Suzuki}}{2,\ast}
}

\address{%
\affil{1,2}{Department of Physics, Kyushu University, 6-10-1 Hakozaki, Higashi-ku, Fukuoka, 812-8581, Japan}
\email{hsuzuki@phys.kyushu-u.ac.jp}
}

\begin{abstract}
In a recent paper~arXiv:1403.4772, we gave a prescription how to construct a
correctly-normalized conserved energy--momentum tensor in lattice gauge theory
containing fermions, on the basis of the Yang--Mills gradient flow. In the
present note, we give an almost identical but somewhat superior prescription
with which one can simply set the fermion mass parameter in our formulation
zero for the massless fermion. This feature will be useful in applying our
formulation to theories in which the masslessness of the fermion is crucial,
such as multi-flavor gauge theories with an infrared fixed point.
\end{abstract}
\subjectindex{B01, B31, B32, B38}
\maketitle

\section{Introduction}
\label{sec:1}
The lattice field theory is incompatible with spacetime symmetries in the
continuum field theory and the construction of the energy--momentum
tensor---the Noether current associated with the translational invariance---is
hence not straightforward. To obtain a correctly-normalized conserved
energy--momentum tensor in the continuum limit, one has to find a linear
combination of (generally Lorentz non-covariant) lattice operators with
non-perturbative coefficients~\cite{Caracciolo:1988hc,Caracciolo:1989pt}; the
coefficients are moreover non-universal in the sense that they depend on the
lattice action adopted. In recent papers~\cite{Suzuki:2013gza,Makino:2014taa},
a completely new approach to construct the energy--momentum tensor on the
lattice has been proposed\footnote{This approach was inspired from a pioneering
experimentation by Itou and~Kitazawa (unpublished).} on the basis of the
ultraviolet (UV) finiteness of the Yang--Mills gradient flow (or the Wilson
flow in the context of lattice gauge
theory)~\cite{Luscher:2010iy,Luscher:2011bx,Luscher:2013cpa}. See
Ref.~\cite{DelDebbio:2013zaa} for a further analysis on this approach. The
formulation of~Ref.~\cite{Suzuki:2013gza} has already been applied to the
thermodynamics of the quenched QCD~\cite{Asakawa:2013laa}.

The pure Yang--Mills theory is treated in~Ref.~\cite{Suzuki:2013gza} and,
in~Ref.~\cite{Makino:2014taa}, a prescription for the vector-like gauge
theories containing massive fermions is given. A somewhat disappointing feature
of the prescription given in~Ref.~\cite{Makino:2014taa} is that since the
fermion variables are normalized by a factor that (to all orders in
perturbation theory) vanishes for the massless fermion, it is not clear whether
one can simply set the mass parameter in the formulation
of~Ref.~\cite{Makino:2014taa} zero for the massless fermion. This might prevent
the application of the formulation to theories in which the masslessness of the
fermion is crucial, such as multi-flavor gauge theories with an infrared fixed
point. The objective of the present note is to remedy this point by giving a
somewhat different prescription which is expected to be free from a possible
singularity associated with the massless fermion.

As in~Ref.~\cite{Makino:2014taa}, we consider an asymptotically-free
vector-like gauge theory with a gauge group~$G$ that contains $N_f$ Dirac
fermions in the gauge representation~$R$. All $N_f$ fermions are assumed to
possess a common mass for simplicity. Since most part of our argument overlaps
with that of~Ref.~\cite{Makino:2014taa}, the basic reasoning and the full
details are referred to~Ref.~\cite{Makino:2014taa}; only essential differences
are presented in the present note. Our notation is identical to that
of~Ref.~\cite{Makino:2014taa}; in particular anti-hermitian generators~$T^a$ of
the representation~$R$ is normalized as~$\tr(T^aT^b)=-T(R)\delta^{ab}$ and the
quadratic Casimir operators are defined by~$T^aT^a=-C_2(R)1$
and~$f^{acd}f^{bcd}=C_2(G)\delta^{ab}$, where $f^{abc}$ are the structure
constants in~$[T^a,T^b]=f^{abc}T^c$. For the fundamental $N$~representation
of~$G=SU(N)$ for which $\dim(N)=N$, we set
\begin{equation}
   C_2(SU(N))=N,\qquad T(N)=\frac{1}{2},\qquad
   C_2(N)=\frac{N^2-1}{2N}.
\label{eq:(1.1)}
\end{equation}

\section{Energy--momentum tensor from the Yang--Mills gradient flow:
A new simpler prescription}

\subsection{Ringed fermion fields}
Local products of gauge fields deformed by the Yang--Mills gradient
flow~\cite{Luscher:2010iy} are UV finite without the multiplicative
renormalization~\cite{Luscher:2011bx} and this property makes the construction
of finite composite operators quite simple. Unfortunately, this finiteness does
not hold for matter fields and the flowed fermion fields\footnote{$t$ denotes
the flow time.} in fact require the multiplicative
renormalization~\cite{Luscher:2013cpa}
\begin{equation}
   \chi_R(t,x)=Z_\chi^{1/2}\chi(t,x),\qquad
   \Bar{\chi}_R(t,x)=Z_\chi^{1/2}\Bar{\chi}(t,x),\qquad
   Z_\chi=1+\frac{g^2}{(4\pi)^2}C_2(R)3\frac{1}{\epsilon}+O(g^4),
\label{eq:(2.1)}
\end{equation}
in the minimal subtraction ($\text{MS}$) scheme, for example. This
renormalization introduces a complication to our problem because one has to
find a matching factor between the multiplicative renormalization in the
dimensional regularization with which the description of the energy--momentum
tensor is simple~\cite{Freedman:1974gs,Joglekar:1975jm} and that in the lattice
regularization. In~Ref.~\cite{Makino:2014taa}, to avoid this complication, we
introduced the ``hatted fermion variables'' by
\begin{equation}
   \Hat{\chi}(t,x)
   =\sqrt{\frac{-2\dim(R)N_fm}
   {(4\pi)^2t\left\langle\Bar{\chi}(t,x)\chi(t,x)\right\rangle}}
   \,\chi(t,x),\qquad
   \Hat{\Bar{\chi}}(t,x)
   =\sqrt{\frac{-2\dim(R)N_fm}
   {(4\pi)^2t\left\langle\Bar{\chi}(t,x)\chi(t,x)\right\rangle}}
   \,\Bar{\chi}(t,x),
\label{eq:(2.2)}
\end{equation}
where $m$ denotes the renormalized fermion mass. Since the multiplicative
renormalization factor~$Z_\chi$ is cancelled out in~$\Hat{\chi}(t,x)$ and
in~$\Hat{\Bar{\chi}}(t,x)$, UV finite composite operators can be constructed by
taking simple local products of $\Hat{\chi}(t,x)$
and~$\Hat{\Bar{\chi}}(t,x)$~\cite{Luscher:2011bx,Luscher:2013cpa}.

The above hatted variables~\eqref{eq:(2.2)} are perfect for massive fermions.
However, since the variables are normalized by the scalar condensation and the
scalar condensation is proportional to the fermion mass at least to all orders
in the perturbation theory, it is not clear for the massless fermion whether
one can simply set the mass parameter in formulas in~Ref.~\cite{Makino:2014taa}
zero without encountering any singular behavior. This point might be cumbersome
for theories in which the masslessness of the fermion is crucial, such as
many-flavor gauge theories with an infrared fixed point (for a recent review,
see Ref.~\cite{Itou:2013faa}).

\begin{figure}
\begin{center}
\includegraphics[width=3cm,clip]{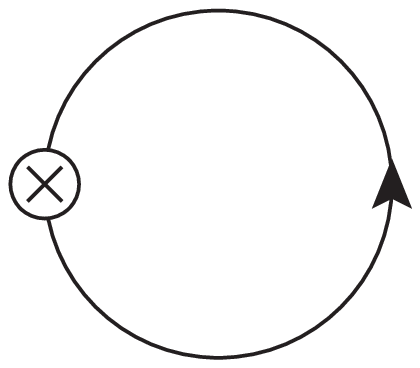}
\caption{D01}
\label{fig:1}
\end{center}
\end{figure}

The proposal we make in the present note is to normalize fermion variables by
using the vacuum expectation value of the fermion kinetic operator, rather than
the scalar condensation. To be precise, we introduce the following ``ringed
variables'',
\begin{align}
   \mathring{\chi}(t,x)
   &=\sqrt{\frac{-2\dim(R)N_f}
   {(4\pi)^2t^2
   \left\langle\Bar{\chi}(t,x)\overleftrightarrow{\Slash{D}}\chi(t,x)
   \right\rangle}}
   \,\chi(t,x),
\label{eq:(2.3)}\\
   \mathring{\Bar{\chi}}(t,x)
   &=\sqrt{\frac{-2\dim(R)N_f}
   {(4\pi)^2t^2
   \left\langle\Bar{\chi}(t,x)\overleftrightarrow{\Slash{D}}\chi(t,x)
   \right\rangle}}
   \,\Bar{\chi}(t,x),
\label{eq:(2.4)}
\end{align}
where
\begin{equation}
   \overleftrightarrow{D}_\mu\equiv D_\mu-\overleftarrow{D}_\mu,\qquad
   D_\mu\equiv\partial_\mu+B_\mu,\qquad
   \overleftarrow{D}_\mu\equiv\overleftarrow{\partial}_\mu-B_\mu,
\label{eq:(2.5)}
\end{equation}
being $B_\mu$ the flowed gauge potential. Again, since the multiplicative
renormalization factor~$Z_\chi$ in~Eq.~\eqref{eq:(2.1)} is cancelled out
in~$\mathring{\chi}(t,x)$ and in~$\mathring{\Bar{\chi}}(t,x)$, UV finite
composite operators can be constructed by simple local products of
$\mathring{\chi}(t,x)$ and~$\mathring{\Bar{\chi}}(t,x)$. Furthermore, those
variables are expected to be non-singular even for the massless fermion. In
fact, to the leading order in the loop expansion (diagram~D01
in~Fig.~\ref{fig:1}; the cross denotes the composite
operator~$\Bar{\chi}(t,x)\overleftrightarrow{\Slash{D}}\chi(t,x)$), we have
\begin{equation}
   \left\langle\Bar{\chi}(t,x)\overleftrightarrow{\Slash{D}}\chi(t,x)
   \right\rangle
   =\frac{-2\dim(R)N_f}{(4\pi)^2t^2}(8\pi t)^\epsilon\left[1+O(m_0^2t)\right],
\label{eq:(2.6)}
\end{equation}
for $D=4-2\epsilon$ dimensions. Since this does not vanish even for the
massless fermion (at least to all orders in the perturbation theory), we expect
that the normalization by this quantity is not singular even for the massless
fermion. Note that the mass dimension which is required for the vacuum
expectation value is supplied by the flow time~$t$ in the present setup.

\begin{figure}
\begin{minipage}{0.3\hsize}
\begin{center}
\includegraphics[width=3cm,clip]{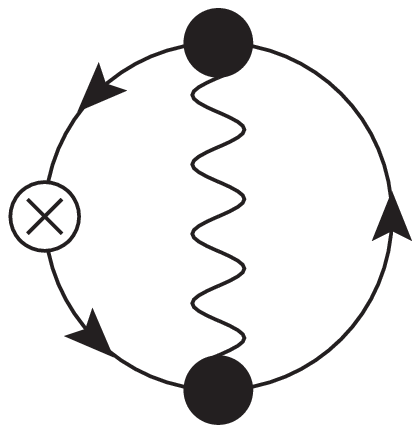}
\caption{D02}
\label{fig:2}
\end{center}
\end{minipage}
\begin{minipage}{0.3\hsize}
\begin{center}
\includegraphics[width=3cm,clip]{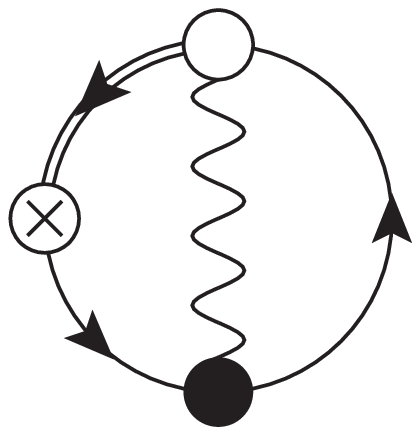}
\caption{D03}
\label{fig:3}
\end{center}
\end{minipage}
\begin{minipage}{0.3\hsize}
\begin{center}
\includegraphics[width=3cm,clip]{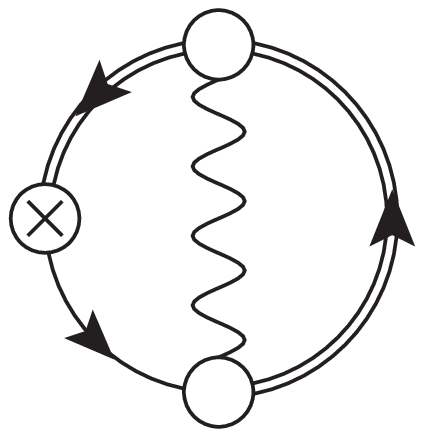}
\caption{D04}
\label{fig:4}
\end{center}
\end{minipage}
\end{figure}

\begin{figure}
\begin{minipage}{0.3\hsize}
\begin{center}
\includegraphics[width=3cm,clip]{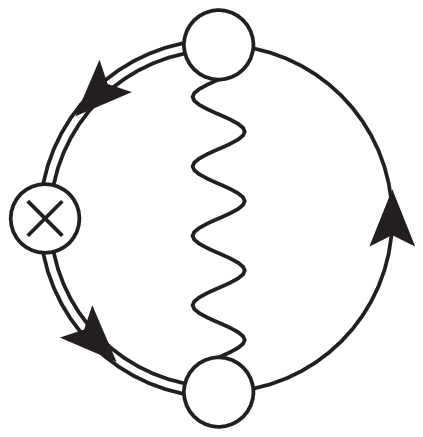}
\caption{D05}
\label{fig:5}
\end{center}
\end{minipage}
\begin{minipage}{0.3\hsize}
\begin{center}
\includegraphics[width=3cm,clip]{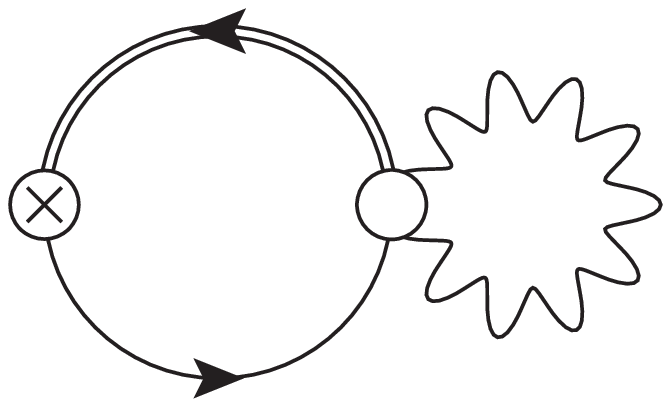}
\caption{D06}
\label{fig:6}
\end{center}
\end{minipage}
\begin{minipage}{0.3\hsize}
\begin{center}
\includegraphics[width=3cm,clip]{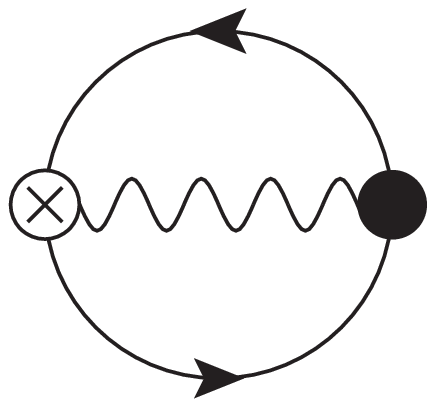}
\caption{D07}
\label{fig:7}
\end{center}
\end{minipage}
\end{figure}

\begin{figure}
\begin{center}
\includegraphics[width=3cm,clip]{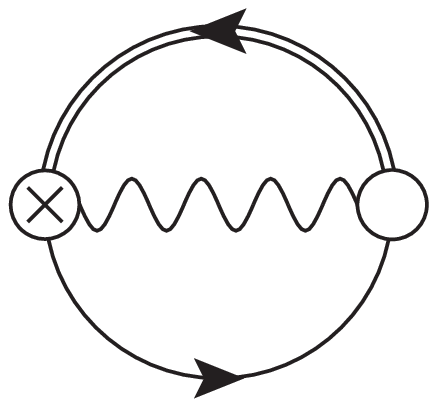}
\caption{D08}
\label{fig:8}
\end{center}
\end{figure}

\begin{table}
\caption{Contribution of each diagram to Eq.~\eqref{eq:(2.7)} in the unit
of~$\frac{-2\dim(R)N_f}{(4\pi)^2t^2}\frac{g_0^2}{(4\pi)^2}C_2(R)$.}
\label{table:1}
\begin{center}
\renewcommand{\arraystretch}{2.2}
\setlength{\tabcolsep}{20pt}
\begin{tabular}{cr}
\toprule
 diagram & \\
\midrule
 D02 & $-\dfrac{1}{\epsilon}-2\ln(8\pi t)+O(m_0^2t)$ \\
 D03 & $2\dfrac{1}{\epsilon}+4\ln(8\pi t)+2+4\ln2-2\ln3+O(m_0^2t)$ \\
 D04 & $-20\ln2+16\ln3+O(m_0^2t)$ \\
 D05 & $12\ln2-5\ln3+O(m_0^2t)$ \\
 D06 & $-4\dfrac{1}{\epsilon}-8\ln(8\pi t)-2+O(m_0^2t)$ \\
 D07 & $8\ln2-4\ln3+O(m_0^2t)$ \\
 D08 & $-2\ln3+O(m_0^2t)$ \\
\bottomrule
\end{tabular}
\end{center}
\end{table}

The next to leading order expression for the expectation
value~\eqref{eq:(2.6)} is given by the flow Feynman diagrams
in~Figs.~\ref{fig:2}--\ref{fig:8} (and diagrams with arrows with the opposite
direction). See Ref.~\cite{Makino:2014taa} for our convention for flow Feynman
diagrams. These diagrams can be evaluated in a similar manner as Appendix~B
of~Ref.~\cite{Luscher:2010iy} and the contribution of each diagram is tabulated
in~Table~\ref{table:1}. Totally, we have
\begin{align}
   &\left\langle\Bar{\chi}(t,x)\overleftrightarrow{\Slash{D}}\chi(t,x)
   \right\rangle
\notag\\
   &=\frac{-2\dim(R)N_f}{(4\pi)^2t^2}
   \left\{(8\pi t)^\epsilon
   +\frac{g_0^2}{(4\pi)^2}C_2(R)
   \left[-3\frac{1}{\epsilon}-6\ln(8\pi t)+\ln(432)\right]
   +O(m_0^2t)+O(g_0^4)\right\}.
\label{eq:(2.7)}
\end{align}
Recalling the bare gauge coupling~$g_0$ and the renormalized gauge coupling~$g$
are related as~$g_0^2=\mu^{2\epsilon}g^2[1+O(g^2)]$, we see that the
normalization factor in~Eqs.~\eqref{eq:(2.3)} and~\eqref{eq:(2.4)} is given by
\begin{equation}
   \frac{-2\dim(R)N_f}
   {(4\pi)^2t^2\left\langle\Bar{\chi}(t,x)
   \overleftrightarrow{\Slash{D}}\chi(t,x)\right\rangle}
   =Z(\epsilon)
   \left\{1+\frac{g^2}{(4\pi)^2}C_2(R)\left[3\frac{1}{\epsilon}-\Phi(t)\right]
   +O(m^2t)+O(g^4)\right\},
\label{eq:(2.8)}
\end{equation}
where
\begin{equation}
   Z(\epsilon)\equiv\frac{1}{(8\pi t)^\epsilon},
\label{eq:(2.9)}
\end{equation}
and
\begin{equation}
   \Phi(t)\equiv-3\ln(8\pi\mu^2t)+\ln(432).
\label{eq:(2.10)}
\end{equation}
Comparison of~Eq.~\eqref{eq:(2.8)} with~Eq.~(3.24)
of~Ref.~\cite{Makino:2014taa} shows that the change from the hatted fermion
variables in~Ref.~\cite{Makino:2014taa} and to the present ringed fermion
variables~\eqref{eq:(2.3)} and~\eqref{eq:(2.4)} entails the following changes
in the expressions of~Ref.~\cite{Makino:2014taa}:
\begin{equation}
   z(\epsilon)\to Z(\epsilon),\qquad
   \phi\to\Phi(t).
\label{eq:(2.11)}
\end{equation}

\subsection{Operator basis and the small flow-time expansion}
To express the energy--momentum tensor in terms of local products of flowed
fields, we introduce following five combinations which are even under the
CP transformation,
\begin{align}
   \Tilde{\mathcal{O}}_{1\mu\nu}(t,x)&\equiv
   G_{\mu\rho}^a(t,x)G_{\nu\rho}^a(t,x),
\label{eq:(2.12)}\\
   \Tilde{\mathcal{O}}_{2\mu\nu}(t,x)&\equiv
   \delta_{\mu\nu}G_{\rho\sigma}^a(t,x)G_{\rho\sigma}^a(t,x),
\label{eq:(2.13)}\\
   \Tilde{\mathcal{O}}_{3\mu\nu}(t,x)&\equiv
   \mathring{\Bar{\chi}}(t,x)
   \left(\gamma_\mu\overleftrightarrow{D}_\nu
   +\gamma_\nu\overleftrightarrow{D}_\mu\right)
   \mathring{\chi}(t,x),
\label{eq:(2.14)}\\
   \Tilde{\mathcal{O}}_{4\mu\nu}(t,x)&\equiv
   \delta_{\mu\nu}
   \mathring{\Bar{\chi}}(t,x)
   \overleftrightarrow{\Slash{D}}
   \mathring{\chi}(t,x),
\label{eq:(2.15)}\\
   \Tilde{\mathcal{O}}_{5\mu\nu}(t,x)&\equiv
   \delta_{\mu\nu}
   m\mathring{\Bar{\chi}}(t,x)
   \mathring{\chi}(t,x),
\label{eq:(2.16)}
\end{align}
where $G_{\mu\nu}^a(t,x)$ is the field strength of the flowed gauge field.
Because of the UV finiteness of the gradient
flow~\cite{Luscher:2011bx,Luscher:2013cpa}, these local products are finite
when expressed in terms of renormalized parameters.

We introduce also corresponding bare operators in the $D$-dimensional
$x$-space:
\begin{align}
   \mathcal{O}_{1\mu\nu}(x)&\equiv
   F_{\mu\rho}^a(x)F_{\nu\rho}^a(x),
\label{eq:(2.17)}\\
   \mathcal{O}_{2\mu\nu}(x)&\equiv
   \delta_{\mu\nu}F_{\rho\sigma}^a(x)F_{\rho\sigma}^a(x),
\label{eq:(2.18)}\\
   \mathcal{O}_{3\mu\nu}(x)&\equiv
   \Bar{\psi}(x)
   \left(\gamma_\mu\overleftrightarrow{D}_\nu
   +\gamma_\nu\overleftrightarrow{D}_\mu\right)
   \psi(x),
\label{eq:(2.19)}\\
   \mathcal{O}_{4\mu\nu}(x)&\equiv
   \delta_{\mu\nu}
   \Bar{\psi}(x)
   \overleftrightarrow{\Slash{D}}
   \psi(x),
\label{eq:(2.20)}\\
   \mathcal{O}_{5\mu\nu}(x)&\equiv
   \delta_{\mu\nu}
   m_0\Bar{\psi}(x)
   \psi(x).
\label{eq:(2.21)}
\end{align}

We then expect that, according to the general argument~\cite{Luscher:2011bx},
the following asymptotic expansion for~$t\to0$ holds
\begin{equation}
   \Tilde{\mathcal{O}}_{i\mu\nu}(t,x)
   =\left\langle\Tilde{\mathcal{O}}_{i\mu\nu}(t,x)\right\rangle
   +\zeta_{ij}(t)
   \left[\mathcal{O}_{j\mu\nu}(x)
   -\left\langle\mathcal{O}_{j\mu\nu}(x)\right\rangle\right]+O(t).
\label{eq:(2.22)}
\end{equation}
Once the mixing coefficients~$\zeta_{ij}(t)$ in this expression are known, one
may invert this relation up to $O(t)$ terms and express the energy-momentum
tensor in the dimensional regularization\footnote{We define the renormalized
finite energy--momentum tensor by subtracting its possible vacuum expectation
value.}
\begin{align}
   \left\{T_{\mu\nu}\right\}_R(x)
   &=\frac{1}{g_0^2}\left\{
   \mathcal{O}_{1\mu\nu}(x)
   -\left\langle\mathcal{O}_{1\mu\nu}(x)\right\rangle
   -\frac{1}{4}
   \left[\mathcal{O}_{2\mu\nu}(x)
   -\left\langle\mathcal{O}_{2\mu\nu}(x)\right\rangle\right]
   \right\}
\notag\\
   &\qquad{}
   +\frac{1}{4}
   \left[\mathcal{O}_{3\mu\nu}(x)
   -\left\langle\mathcal{O}_{3\mu\nu}(x)\right\rangle\right]
   -\frac{1}{2}
   \left[\mathcal{O}_{4\mu\nu}(x)
   -\left\langle\mathcal{O}_{4\mu\nu}(x)\right\rangle\right]
\notag\\
   &\qquad\qquad{}
   -\left[\mathcal{O}_{5\mu\nu}(x)
   -\left\langle\mathcal{O}_{5\mu\nu}(x)\right\rangle\right],
\label{eq:(2.23)}
\end{align}
in terms of the local products in~Eqs.~\eqref{eq:(2.12)}--\eqref{eq:(2.16)}.
The resulting expression will be\footnote{In the first line, we have used the
fact that the finite operator $\Tilde{\mathcal{O}}_{1\mu\nu}(t,x)
-(1/4)\Tilde{\mathcal{O}}_{2\mu\nu}(t,x)$ is traceless in~$D=4$ and thus has no
vacuum expectation value.}
\begin{align}
   \left\{T_{\mu\nu}\right\}_R(x)
   &=c_1(t)\left[
   \Tilde{\mathcal{O}}_{1\mu\nu}(t,x)
   -\frac{1}{4}\Tilde{\mathcal{O}}_{2\mu\nu}(t,x)
   \right]
\notag\\
   &\qquad{}
   +c_2(t)\left[
   \Tilde{\mathcal{O}}_{2\mu\nu}(t,x)
   -\left\langle\Tilde{\mathcal{O}}_{2\mu\nu}(t,x)\right\rangle
   \right]
\notag\\
   &\qquad\qquad{}
   +c_3(t)\left[
   \Tilde{\mathcal{O}}_{3\mu\nu}(t,x)
   -2\Tilde{\mathcal{O}}_{4\mu\nu}(t,x)
   -\left\langle
   \Tilde{\mathcal{O}}_{3\mu\nu}(t,x)
   -2\Tilde{\mathcal{O}}_{4\mu\nu}(t,x)
   \right\rangle
   \right]
\notag\\
   &\qquad\qquad\qquad{}
   +c_4(t)\left[
   \Tilde{\mathcal{O}}_{4\mu\nu}(t,x)
   -\left\langle\Tilde{\mathcal{O}}_{4\mu\nu}(t,x)\right\rangle
   \right]
\notag\\
   &\qquad\qquad\qquad\qquad{}
   +c_5(t)\left[
   \Tilde{\mathcal{O}}_{5\mu\nu}(t,x)
   -\left\langle\Tilde{\mathcal{O}}_{5\mu\nu}(t,x)\right\rangle
   \right]+O(t).
\label{eq:(2.24)}
\end{align}

Eq.~\eqref{eq:(2.24)} shows that the energy--momentum tensor can be obtained as
the $t\to0$ limit of the combination in the right-hand side. Since the UV
finite composite operators~\eqref{eq:(2.12)}--\eqref{eq:(2.16)} should be
independent of the regularization adopted, one may use the lattice
regularization to compute correlation functions of the quantity in the
right-hand side of~Eq.~\eqref{eq:(2.24)}. This provides a possible method to
compute correlation functions of the correctly-normalized conserved
energy--momentum tensor with the lattice
regularization~\cite{Suzuki:2013gza,Makino:2014taa}.

In Ref.~\cite{Makino:2014taa}, by using the hatted variables~\eqref{eq:(2.2)},
the mixing coefficients~$\zeta_{ij}(t)$ in~Eq.~\eqref{eq:(2.22)} and the
corresponding coefficients~$c_i(t)$ in~Eq.~\eqref{eq:(2.24)} were computed in
the perturbation theory to the one-loop order. This perturbative computation is
justified for~$t\to0$ by a renormalization group argument (see below).
Fortunately, we do not need to repeat this computation anew for the present
ringed variables~\eqref{eq:(2.3)} and~\eqref{eq:(2.4)}; the only difference is
the normalization of the fermion variables which amounts to the changes
in~Eq.~\eqref{eq:(2.11)}. Making these changes in~Eqs.~(4.62)--(4.66)
of~Ref.~\cite{Makino:2014taa}, we have
\begin{align}
   c_1(t)&
   =\frac{1}{g^2}-b_0\ln(8\pi\mu^2t)-\frac{7}{8}\frac{1}{(4\pi)^2}
   \left[\frac{11}{3}C_2(G)
   -\frac{12}{7}T(R)N_f\right],
\label{eq:(2.25)}\\
   c_2(t)&
   =\frac{1}{8}
   \frac{1}{(4\pi)^2}
   \left[\frac{11}{3}C_2(G)
   +\frac{11}{3}T(R)N_f\right],
\label{eq:(2.26)}\\
   c_3(t)&
   =\frac{1}{4}\left\{1+\frac{g^2}{(4\pi)^2}C_2(R)
   \left[\frac{3}{2}+\ln(432)\right]\right\},
\label{eq:(2.27)}\\
   c_4(t)&=\frac{1}{8}d_0g^2,
\label{eq:(2.28)}\\
   c_5(t)&=-\left\{1+\frac{g^2}{(4\pi)^2}C_2(R)
   \left[3\ln(8\pi\mu^2t)+\frac{7}{2}+\ln(432)\right]\right\},
\label{eq:(2.29)}
\end{align}
where the $\text{MS}$ scheme is assumed and
\begin{equation}
   b_0=\frac{1}{(4\pi)^2}
   \left[\frac{11}{3}C_2(G)-\frac{4}{3}T(R)N_f\right],\qquad
   d_0=\frac{1}{(4\pi)^2}6C_2(R).
\label{eq:(2.30)}
\end{equation}

\subsection{Renormalization group argument}
The use of the perturbation theory in computing $c_i(t)$ for~$t\to0$ is
justified by the renormalization group argument. We apply
\begin{equation}
   \left(\mu\frac{\partial}{\partial\mu}\right)_0,
\label{eq:(2.31)}
\end{equation}
to both sides of~Eq.~\eqref{eq:(2.24)}, where $\mu$ is the renormalization
scale and the subscript~$0$ implies that the derivative is taken while all bare
quantities are kept fixed. Since the energy--momentum tensor is not
multiplicatively renormalized,
$(\mu\partial/\partial\mu)_0(\text{left-hand side of Eq.~\eqref{eq:(2.24)}})=0$.
On the right-hand side, since $\Tilde{\mathcal{O}}_{1,2,3,4\mu\nu}(t,x)$
and~$(1/m)\Tilde{\mathcal{O}}_{5\mu\nu}(t,x)$
in~Eqs.~\eqref{eq:(2.12)}--\eqref{eq:(2.16)} are entirely given by bare
quantities through the flow equations
in~Refs.~\cite{Luscher:2010iy,Luscher:2011bx,Luscher:2013cpa}, we have
\begin{equation}
   \left(\mu\frac{\partial}{\partial\mu}\right)_0
   \Tilde{\mathcal{O}}_{1,2,3,4\mu\nu}(t,x)
   =\left(\mu\frac{\partial}{\partial\mu}\right)_0
   \frac{1}{m}\Tilde{\mathcal{O}}_{5\mu\nu}(t,x)
   =0.
\label{eq:(2.32)}
\end{equation}
These observations imply,
\begin{equation}
   \left(\mu\frac{\partial}{\partial\mu}\right)_0c_{1,2,3,4}(t)
   =\left(\mu\frac{\partial}{\partial\mu}\right)_0mc_{5}(t)
   =0.
\label{eq:(2.33)}
\end{equation}
Then the standard renormalization group argument tells that $c_{1,2,3,4}(t)$
and~$mc_5(t)$ are independent of the renormalization scale, if the renormalized
parameters in these quantities are replaced by running parameters defined by
\begin{align}
   &q\frac{d\Bar{g}(q)}{dq}=\beta(\Bar{g}(q)),\qquad
   \Bar{g}(q=\mu)=g,
\label{eq:(2.34)}\\
   &q\frac{d\Bar{m}(q)}{dq}=-\gamma_m(\Bar{g}(q))\Bar{m}(q),\qquad
   \Bar{m}(q=\mu)=m,
\label{eq:(2.35)}
\end{align}
where $\mu$ is the original renormalization scale. Thus, since $c_{1,2,3,4}(t)$
and~$mc_5(t)$ are independent of the renormalization scale, two possible
choices, $q=\mu$ and~$q=1/\sqrt{8t}$, should give an identical result. In this
way, we have
\begin{align}
   c_{1,2,3,4}(t)(g,m;\mu)
   &=c_{1,2,3,4}(t)(\Bar{g}(1/\sqrt{8t}),\Bar{m}(1/\sqrt{8t});1/\sqrt{8t}),
\label{eq:(2.36)}\\
   c_5(t)(g,m;\mu)
   &=\frac{\Bar{m}(1/\sqrt{8t})}{m}
   c_5(t)(\Bar{g}(1/\sqrt{8t}),\Bar{m}(1/\sqrt{8t});1/\sqrt{8t}),
\label{eq:(2.37)}
\end{align}
where we have explicitly written dependence of $c_i(t)$ on renormalized
parameters and on the renormalization scale. Finally, since the running gauge
coupling $\Bar{g}(1/\sqrt{8t})\to0$ for~$t\to0$ thanks to the asymptotic
freedom, we infer that we can compute $c_i(t)$ for~$t\to0$ by using the
perturbation theory.

Applying Eqs.~\eqref{eq:(2.36)} and~\eqref{eq:(2.37)} to
Eqs.~\eqref{eq:(2.25)}--\eqref{eq:(2.29)}, we have
\begin{align}
   c_1(t)&
   =\frac{1}{\Bar{g}(1/\sqrt{8t})^2}
   -b_0\ln\pi-\frac{7}{8}\frac{1}{(4\pi)^2}
   \left[\frac{11}{3}C_2(G)
   -\frac{12}{7}T(R)N_f\right],
\label{eq:(2.38)}\\
   c_2(t)&
   =\frac{1}{8}
   \frac{1}{(4\pi)^2}
   \left[\frac{11}{3}C_2(G)
   +\frac{11}{3}T(R)N_f\right],
\label{eq:(2.39)}\\
   c_3(t)&
   =\frac{1}{4}\left\{1+\frac{\Bar{g}(1/\sqrt{8t})^2}{(4\pi)^2}C_2(R)
   \left[\frac{3}{2}+\ln(432)\right]\right\},
\label{eq:(2.40)}\\
   c_4(t)&=\frac{1}{8}d_0\Bar{g}(1/\sqrt{8t})^2,
\label{eq:(2.41)}\\
   c_5(t)&=-\frac{\Bar{m}(1/\sqrt{8t})}{m}
   \left\{1+\frac{\Bar{g}(1/\sqrt{8t})^2}{(4\pi)^2}C_2(R)
   \left[3\ln\pi+\frac{7}{2}+\ln(432)\right]\right\}.
\label{eq:(2.42)}
\end{align}
The above expressions are for the $\text{MS}$ scheme. The expressions in the
$\overline{\text{MS}}$ scheme can be obtained by making the replacement
\begin{equation}
   \ln\pi\to\gamma_E-2\ln2,
\label{eq:(2.43)}
\end{equation}
in the above expressions.

This completes our construction of the lattice energy--momentum tensor in a
new simpler prescription, which is expected to be free from a possible
singularity accosiated with the massless fermion. The energy--momentum tensor
is given by the $t\to0$ limit of the right-hand side of~Eq.~\eqref{eq:(2.24)},
where the coefficients~$c_i(t)$ are given
by~Eqs.~\eqref{eq:(2.38)}--\eqref{eq:(2.42)}; for the massless fermion, we can
simply set $c_5(t)=0$ or discard the
operator~$\Tilde{\mathcal{O}}_{5\mu\nu}(t,x)$. One may use any lattice
transcription of operators~\eqref{eq:(2.12)}--\eqref{eq:(2.16)} because these
operators are UV finite. Note that
coefficients~\eqref{eq:(2.38)}--\eqref{eq:(2.42)} are universal in the sense
that they are common for any lattice action and for any lattice transcription
of operators~\eqref{eq:(2.12)}--\eqref{eq:(2.16)} (as far as the classical
continuum limit of the flow equations are identical to those
of~Refs.~\cite{Luscher:2010iy,Luscher:2011bx,Luscher:2013cpa}). To utilize this
``universality'', however, one has to take the continuum limit before the
$t\to0$ limit for~Eq.~\eqref{eq:(2.24)}. Practically, with a finite lattice
spacing~$a$, the flow time~$t$ cannot be taken arbitrarily small because of a
natural constraint,
\begin{equation}
   a\ll\sqrt{8t}\ll R,
\label{eq:(2.44)}
\end{equation}
where~$R$ denotes a typical low-energy scale. The extrapolation for~$t\to0$
thus generally requires a sufficiently fine lattice. The application to the
thermodynamics of the quenched QCD~\cite{Asakawa:2013laa} strongly indicates
that a reliable extrapolation to $t\to0$ is feasible even with
presently-available lattice parameters.

The work of H.~S. is supported in part by a Grant-in-Aid for Scientific
Research~23540330.


\begin{thebibliography}{00}

\bibitem{Caracciolo:1988hc} 
  S.~Caracciolo, G.~Curci, P.~Menotti and A.~Pelissetto,
  Nucl.\ Phys.\ B {\bf 309}, 612 (1988).

\bibitem{Caracciolo:1989pt} 
  S.~Caracciolo, G.~Curci, P.~Menotti and A.~Pelissetto,
  Annals Phys.\  {\bf 197}, 119 (1990).

\bibitem{Suzuki:2013gza} 
  H.~Suzuki,
  PTEP {\bf 2013}, no. 8, 083B03 (2013)
  [arXiv:1304.0533 [hep-lat]].

\bibitem{Makino:2014taa} 
  H.~Makino and H.~Suzuki,
  arXiv:1403.4772 [hep-lat].

\bibitem{Luscher:2010iy} 
  M.~L\"uscher,
  JHEP {\bf 1008}, 071 (2010)
  [arXiv:1006.4518 [hep-lat]].

\bibitem{Luscher:2011bx} 
  M.~L\"uscher and P.~Weisz,
  JHEP {\bf 1102}, 051 (2011)
  [arXiv:1101.0963 [hep-th]].

\bibitem{Luscher:2013cpa} 
  M.~L\"uscher,
  JHEP {\bf 1304}, 123 (2013)
  [arXiv:1302.5246 [hep-lat]].

\bibitem{DelDebbio:2013zaa} 
  L.~Del Debbio, A.~Patella and A.~Rago,
  JHEP {\bf 1311}, 212 (2013)
  [arXiv:1306.1173 [hep-th]].

\bibitem{Asakawa:2013laa} 
  M.~Asakawa {\it et al.}  [FlowQCD Collaboration],
  arXiv:1312.7492 [hep-lat].

\bibitem{Freedman:1974gs} 
  D.~Z.~Freedman, I.~J.~Muzinich and E.~J.~Weinberg,
  Annals Phys.\  {\bf 87}, 95 (1974).

\bibitem{Joglekar:1975jm} 
  S.~D.~Joglekar,
  Annals Phys.\  {\bf 100}, 395 (1976)
  [Erratum-ibid.\  {\bf 102}, 594 (1976)].

\bibitem{Itou:2013faa} 
  E.~Itou,
  arXiv:1311.2676 [hep-lat].

\end{thebibliography}
\end{document}